\pdfoutput=1

\documentclass[floatfix,superscriptaddress,twocolumn,showpacs,aps,amsmath,amssymb]{revtex4-1}
\usepackage{graphicx}
\usepackage{amssymb}
\usepackage{bm}

\def\tc{$T_{\mathrm{c}}$\ }
\def\tcrho{$T_{\mathrm{c\rho}}$\ }
\def\tcchi{$T_{\mathrm{c\chi}}$\ }
\def\t6as{$\mathrm{(TMTSF)_{2}AsF_{6}}$\ }
\def\tmx{(TMTSF)$_{2}$(ClO$_{4}$)$_{(1-x)}$(ReO$_{4}$)$_x$\ }

\def\tmc{$\mathrm{(TMTSF)_{2}ClO_{4}}$}
\def\tmcns{$\mathrm{(TMTSF)_{2}ClO_{4}}$}
\def\tms{$\mathrm{(TMTSF)_{2}AsF_{6(1-x)}SbF_{6x}}$\ }

\def\tmttfsbf6{$\mathrm{(TMTTF)_{2}SbF_{6}}$\ }

\def\tmttfasf6{$\mathrm{(TMTTF)_{2}AsF_{6}}$\ }
\def\tmtsfasf6{$\mathrm{(TMTSF)_{2}AsF_{6}}$\ }
\def\tmttfbf4{$\mathrm{(TMTTF)_{2}BF_{4}}$\ }

\def\tmtsfreo4{$\mathrm{(TMTSF)_{2}ReO_{4}}$\ }
\def\tmno3{$\mathrm{(TMTSF)_{2}NO_{3}}$\ }
\def\tm2x{$\mathrm{(TM)_{2}X}$\ }
\def\tm2xns{$\mathrm{(TM)_{2}X}$}

\def\hc2{$H_{\mathrm{c2}}$\ }

\def\tsm{$\mathrm{TMTSF}$\ }

\def\tmp6{$\mathrm{(TMTSF)_{2}PF_{6}}$\ }
\def\tmpns{$\mathrm{(TMTSF)_{2}PF_{6}}$}
\def\tms2x{$\mathrm{(TMTSF)_{2}}X$}
\def\tm2x{$\mathrm{(TM)_{2}}X$\ }

\def\re{$\mathrm{ReO_{4}}$}

\def\cl{$\mathrm{ClO_{4}}$}

\def\4fb{$\mathrm{BF_{4}}$}

\def\reo4{$\mathrm{ReO_{4}}$}
\def\bedtttfreo4{$\mathrm{(BEDT-TTF)_{2}ReO_{4}}$\ }
\def\et2i3{$\mathrm{(ET)_{2}I_{3}}$\,}
\def\et2x{$\mathrm{(ET)_{2}X}$\,}
\def\ket2x{$\mathrm{\kappa-(ET)_{2}X}$\,}

\def\ket2x{$\mathrm{\kappa-(ET)_{2}X}$\,}

\def\betsfecl4{$\mathrm{(BETS)_{2}FeCl_{4}}$\,}

\def\et{$\mathrm{ET}$\,}

\newcommand{\sub}[1]{$_{\mathrm {#1}}$}
\newcommand{\subm}[1]{_{\mathrm {#1}}}
\newcommand{\sps}[1]{$^{\mathrm {#1}}$}
\newcommand{\spsm}[1]{^{\mathrm {#1}}}

\newcommand{\Tc}{T\subm{c}}
\newcommand{\Ta}{T\subm{AO}}

\newcommand{\Tcrz}{T_{\mathrm{c},\rho=0}}
\newcommand{\Tcc}{T_{\mathrm{c}\chi}}
\newcommand{\Tcr}{T_{\mathrm{c}\rho}}

\newcommand{\dash}{^{\prime}}
\newcommand{\ddash}{^{\prime\prime}}

\newcommand{\cstar}{c^{\ast}}

\newcommand{\chire}{\varDelta\chi\subm{AC}\dash}
\newcommand{\chiim}{\varDelta\chi\subm{AC}\ddash}
\newcommand{\rhoc}{\rho_{\cstar}}
\newcommand{\rhocz}{\rho_{\cstar 0}}
\newcommand{\rhoczc}{\rho_{\cstar 0\mathrm{c}}}
\newcommand{\rhoMin}{\rho_{\mathrm{Min}}}
\newcommand{\rhoMax}{\rho_{\mathrm{Max}}}
\newcommand{\VF}{v_{\mathrm{shield}}}
\newcommand{\VX}{v_{\mathrm{XRD}}}
\newcommand{\Tcz}{T_{\mathrm{c0}}}
\newcommand{\Ohmcm}{\Omega\text{\textperiodcentered cm}}

\begin{document}
\title{Crossover from impurity-controlled to granular superconductivity in (TMTSF)$_{\bm{2}}$ClO$_{\bm{4}}$}

\author{Shingo~Yonezawa}\email{yonezawa@scphys.kyoto-u.ac.jp}
\affiliation{Department of Physics, Graduate School of Science,  Kyoto University, Kyoto 606-8502, Japan}

\author{Claire A Marrache-Kikuchi} \affiliation{CSNSM, Univ. Paris-Sud, CNRS/IN2P3, 91405 Orsay, France}

\author{Klaus~Bechgaard} \affiliation{Department~of~Chemistry, Oersted~Institute, Universitetsparken 5, 2100 Copenhagen, Denmark}

\author{Denis~J\'erome} \affiliation{Laboratoire de Physique des Solides (UMR 8502), Univ. Paris-Sud, 91405 Orsay, France}

\date{\today}

\begin{abstract}
 Using a proper cooling procedure, a controllable amount of non-magnetic structural disorder can be introduced at low temperature  in  \tmcns.
Here we performed simultaneous measurements of transport and magnetic properties of \tmcns\ in its normal and superconducting states, while finely covering three orders of magnitude of the cooling rate around the anion ordering temperature. 
Our result reveals, with increasing density of disorder, the existence of a crossover between homogeneous defect-controlled $d$-wave superconductivity and granular superconductivity.  
At slow cooling rates, with small amount of disorder, the evolution of superconducting properties is well described with the Abrikosov-Gorkov theory, providing further confirmation of non-$s$-wave pairing in this compound.
In contrast, at fast cooling rates, zero resistance and diamagnetic shielding are achieved through a randomly distributed network of superconducting puddles embedded in an normal conducting background and interconnected by proximity effect coupling. 
The temperature dependence of the AC complex susceptibility reveals features typical for a network of granular superconductors.
This makes \tmc\ a model system for granular superconductivity where the grain size and their concentration are tunable within the same sample.
\end{abstract}

\maketitle

\section{Introduction}

Most of the recently discovered so-called ``unconventional superconductors'', such as the quasi-one-dimensional (Q1D) organic superconductors~\cite{Jerome80,Bechgaard81}, copper oxides~\cite{Bednorz86}, two-dimensional organic superconductors based on bis-ethylenedithio-tetrathiafulvalene (BEDT-TTF) or related molecules~\cite{McKenzie97,Kanoda97,Lefebvre00}, and the layered iron pnictides~\cite{Kamihara08}, share a unifying property:
to exhibit a general phase diagram where superconductivity has a common border with a magnetically ordered phase.
For such superconductors, the stability of superconductivity against non-magnetic defects provides important information toward understanding of superconducting (SC) properties such as the pairing symmetry.
As a particular example, in the ruthenate superconductor Sr\sub{2}RuO\sub{4}~\cite{Maeno1994}, strong reduction of $\Tc$ by a small amount of non-magnetic impurity lend support for triplet $p$-wave pairing~\cite{Mackenzie1998.PhysRevLett.80.161}.
Since the non-magnetic impurity effect is still actively debated for systems such as multi-band superconductors~\cite{Hirschfeld2016.CRPhys.17.197} and topological superconductors~\cite{Michaeli2012.PhysRevLett.109.187003,Nagai2015.PhysRevB.91.060502}, a detailed and controlled investigation of the impurity effect is an important subject.

Q1D organic superconductors are textbook examples unconventional superconductivity, with the competition between density wave (insulating) and pairing (superconducting) orderings, namely the Peierls and Cooper divergences~\cite{Bychkov66,Emery79,Solyom79}.
Reviews can be found in Refs.~\cite{Jerome82,Ishiguro98,Giamarchi04}.
In the case of \tmpns, a prototype of the \tms2x family, where \tsm is
the tetramethyl-tetrafulvalene electron-donor molecule and $X$ a mono-valent anion, the balance between the two possible ground states is controlled by the magnitude of the kinetic coupling between molecular stacks~\cite{Yamaji83,Montambaux88}.
Hence, the interplay between Peierls and Cooper channels can be controlled by an applied hydrostatic pressure: Under low pressures, a SDW ground state is stable~\cite{Jerome80,Jerome82} due to the good nesting of the warped Fermi surfaces, whereas superconductivity is stable above $\approx 9$~kbar when anti-nesting terms become dominant in their dispersion relation.

The only ambient-pressure superconductor in the \tms2x\ family, namely \tmcns, was found shortly after the discovery of \tmpns~\cite{Bechgaard81}.
The SC order parameter of the \tms2x\ family has been thoroughly investigated mainly on \tmcns~\cite{Takigawa1987,Belin1997,Joo04,Joo05,Shinagawa2007,Yonezawa2012.PhysRevB.85.140502R,Pratt2013.PhysRevLett.110.107005}.
Although the debate is not fully settled yet, one of the most plausible scenarios is the magnetic-fluctuation driven nodal $d$-wave-like pairing, as strongly demonstrated by the field-angle-dependent heat capacity measurements~\cite{Yonezawa2012.PhysRevB.85.140502R,Jerome16}.
Such a possibility has been indeed theoretically proposed even from early days~\cite{Emery86,Hasegawa1986.JPhysSocJpn.55.3978,Bourbonnais86,Hasegawa1987.JPhysSocJpn.56.877,Bourbonnais88a}.
We note that, within this scenario, the gap nodes are accompanied by non-trivial topological features, as in the case of the nodes in $d$-wave superconductors~\cite{Schnyder2015.JPhysCondensMatter.27.243201}. 
Thus \tms2x is a candidate for topological nodal superconductors~\cite{Yonezawa2016.AAPPSBulletin.26.3}, which are of interest in the context of recently developing topological materials science.

The impurity effect on the \tms2x family has been investigated for a long time.
The remarkable sensitivity of superconductivity to irradiation~\cite{Bouffard82,Choi82} has first been considered as a signature for triplet pairing~\cite{Abrikosov83}.
However, defects introduced  by irradiation in a controlled way~\cite{Zuppirolli87} can often be magnetic~\cite{Sanquer85}.
Hence, the suppression of superconductivity by irradiation-induced defects must be taken
with care since local magnetic impurities can act as strong pair-breakers even on $s$-wave superconductors.
A softer way to introduce non-magnetic impurities in the \tms2x series is to produce a solid solution of isoelectronic anions. The solid solution procedure of \tmcns, replacing with the tetrahedral anion ClO\sub{4} with ReO\sub{4}, leads to a suppression of \tc due to genuine non-magnetic impurities~\cite{Tomic83a,Joo04}.
These results provided strong evidence for the realization of a non-$s$-wave pairing state in \tms2x.

The ambient-pressure superconductor \tmcns\ provides another unique means to control its unconventional superconductivity by disorder.
This material exhibits superconductivity below $\Tc \sim 1.3$~K if the sample is cooled slowly~\cite{Bechgaard81,Garoche82, Schwenk82}.
However, if the sample is cooled fast enough in the vicinity of 25~K, \tmc\ undergoes a spin density wave (SDW) transition towards an insulating ground state below 4-5~K~\cite{Tomic82,Ishiguro83,Takahashi83}.
This significant feature of \tmc\ is actually related to the non-centrosymmetric nature of the tetrahedral \cl\ anion located on the inversion centers of the full structure.
At high temperature, thermal motion of the \cl\ orientation makes it possible to preserve inversion symmetry on average since \cl\ occupy randomly one or the other inversion-symmetry-related orientations.
The structural disorder between two possible orientations no longer persists at low temperatures below $\Ta = 24$~K, because an entropy gain due to the reduction of degrees of freedom triggers the anion ordering below this temperature.

The ordering transition has been studied by several techniques.
In particular, diffuse X-ray work has shown that, while \cl\, anions adopt a uniform orientation along $a$ and $c$ axes, they alternate along $b$~\cite{Pouget83b} when the crystal is cooled very slowly through $\Ta$, to reach the so-called relaxed state~\cite{Tomic82,Gubser82}.
Since the dynamics of the anion orientation was shown to be slow~\cite{Pouget12,Tomic82,Schwenk84}, fast cooling (i.e., cooling much faster than 17~K/min by quenching) enables to retain anion disorder even below $\Ta$ in a metastable state with a single pair of Fermi surface sheets resembling closely that of \tmpns.
Consequently, the good nesting of the single-pair Fermi surfaces at $\pm k\subm{F}$ for the quenched material stabilizes the insulating SDW state.
In contrast, with the slow enough cooling rate, the alternating order of the anions along the $b'$ axis leads to a folding of the Fermi surface, opening a gap at $\pm b^{\ast}/4$ and resulting in doubling of the Fermi-surface sheets.
This folding suppresses the SDW state by disturbing the Fermi-surface nesting, instead allowing superconductivity to appear below 1.2-1.3~K.

Although the two extreme situations, namely the relaxed and quenched states, with SC and SDW ground states respectively, have been fairly extensively studied from the early days, only limited studies have been published for the intermediate cooling-rate regime, where superconductivity is moderately suppressed~\cite{Garoche82a,Matsunaga99,Joo04}.
High-resolution X-ray investigations~\cite{Kagoshima83,Moret85,Pouget90} have shown that samples in the intermediate cooling-rate regime exhibit a peculiar anion ordering, in which domains with ordered anions of finite size are embedded in a disordered background.
In the regime up to 5~K/min, high resolution X-ray diffraction measurements have determined both the volume fraction of ordered anions and the average size of ordered domains~\cite{Pouget90}.

Here, we report on the resistivity $\rhoc$ along the least conducting $\cstar$ axis~\cite{NoteRhoC} simultaneously measured with the AC susceptibility $\chi\subm{AC}$ under carefully controlled cooling across $\Ta$, in order to address the changes of properties of the normal and SC phases by the increase of disorder.
With the experimental settings especially designed for the present investigation, we explored the cooling rate around $\Ta$ from 0.02 to 18~K/min with fine steps, nearly covering three orders of magnitude.
We believe that the present study is one of the most careful impurity-effect studies in any unconventional superconductors, because we  control the amount of the defects in identical samples, by a reversible way, and with very fine intervals.
From our transport and magnetic measurements, we reveal that the intermediate cooling-rate regime of \tmc\ cannot be considered as a state exhibiting an average anion order/disorder evolving according to the cooling rate, but instead as a state comprising a cooling rate-dependent volume fraction of well-ordered domains with the rest of the volume occupied by disordered anions. 
We also study the evolution of the onset $\Tc$ of $\chi\subm{AC}(T)$ and that of $\rhoc(T)$.
We note that $\Tc$ under fast cooling has been studied using a time-dependent Ginzburg-Landau theory~\cite{Haddad11}. 
However, this earlier study was only based on transport properties.

\section{Experimental}

For the present study, two single crystals of \tmc from the same batch grown with an electrocrystallization method have been used.
Simultaneous transport and susceptibility measurements have been performed on one sample (\#1; with the size of $\approx 2.4\times 0.7 \times 0.1$~mm$^3$).
The results thus obtained will constitute the core of the paper.
With a slightly different set-up with an additional thermometer directly attached beneath the sample, the AC susceptibility was measured for another sample (\#2; with the size of $\approx 1.5\times 0.4 \times 0.1$~mm$^3$).
Results for Sample \#2 and details of the measurement setup will be given in the Supplementary Materials~\cite{NoteSupple}.

For both experiments, the samples were cooled down with a commercial cryostat (Quantum Design; PPMS) with an adiabatic demagnetization refrigerator (ADR) option.
In order to control the disorder in the ClO\sub{4} orientation within a sample, we first raised the temperature to 50~K and kept it for more than one hour to fully randomize the anion orientations.
The sample was first cooled in a controlled way down to 10~K using the temperature controller of the PPMS.
After further cooling down to 1.8~K, the temperature was kept at 1.8~K and the DC magnetic field of 2.5~T applied.
The sample chamber was then evacuated with a turbo molecular pump to achieve high-vacuum adiabatic condition.
Then the magnetic field was turned off to reach $\sim 0.1$~K, and the transport or magnetic data were collected during warming.
See Fig.~S2 of the Supplementary Materials~\cite{NoteSupple} as well as Fig.~8 of Ref.~\cite{Yonezawa15} for details of the cooling sequence of the ADR option.
We should comment here that the sample was always in the field-cooled condition although all measurements were performed in zero DC field.

\begin{figure}[tbh]
\begin{center}
\includegraphics[width=9cm]{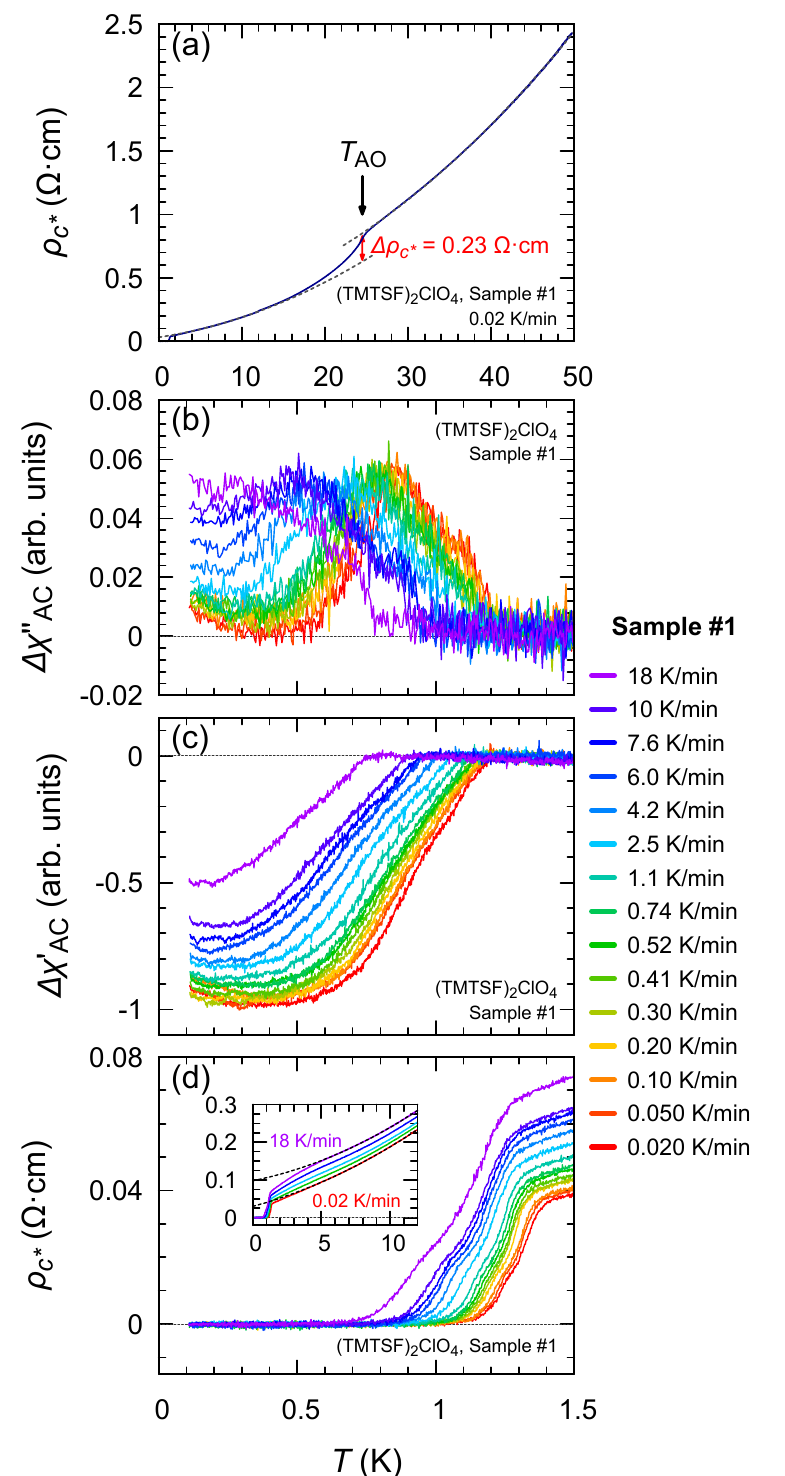}
\caption{(color online)
Normal and superconducting properties of \tmc for Sample~\#1 at various cooling rates.
(a) Inter-layer resistivity $\rhoc$ versus $T$ at 0.020~K/min showing the drop of the elastic resistivity $\Delta\rhoc = 0.23~\Ohmcm$ at $\Ta$ as indicated by the red arrow.
(b) Change in the imaginary part of the AC susceptibility.
(c) Change in the real part of the AC susceptibility.
(d) $\rhoc(T)$ simultaneously measured with the data in (b) and (c).
The inset presents $\rhoc(T)$ on a broader temperature range for 0.020, 0.52, 2.5, 7.6, and 18~K/min.
For the 0.020 and 18-K/min data, the results of the fitting, performed using the procedure explained in the text, are presented with the broken curves.
\label{fig:raw}}
\end{center}
\end{figure}
\label{cooling rate dependence}

The resistivity of the sample were measured with the resistivity option of PPMS.
The current of 10~$\mu$A (1~mA for measurements at high temperatures) along the $\cstar$-axis direction was reversed to cancel the thermoelectric voltage.
The AC susceptibility was measured with an AC field of  $\sim 0.17$~Oe-rms at 3011~Hz (Sample \#1) or 887~Hz (Sample \#2) parallel to the $\cstar$ axis, using a conventional mutual inductance method with a miniature susceptometer~\cite{Yonezawa15}
fitting into the ADR option of the PPMS.
The output signal of the susceptometer was measured using a lock-in amplifier (Stanford Research Systems, SR830).
To correct a small mixing of the real and imaginary parts in the lock-in amplifier signal, an additional phase factor $\theta$ is adopted: $\chire \propto \Delta V_y\cos(\theta) + \Delta V_x\sin(\theta)$ and $\chiim \propto \Delta V_x\cos(\theta) - \Delta V_y\sin(\theta)$, where $\Delta V_x$ and $\Delta V_y$ are changes in the reading of the lock-in amplifier. The value of $\theta$ is chosen as $\theta = 1.5^\circ$ and $8.5^\circ$ for Samples~\#1 and \#2, respectively, so that $\chiim =0$ for $T>\Tc$ and $T\to 0$ for the 0.020~K/min data.

\section{Experimental data}

Figure \ref{fig:raw}(a) displays temperature dependence of $\rhoc$ up to 50~K upon cooling at 0.02~K/min.
A clear anomaly at $\Ta=24.5$~K is observed.
By comparing the fittings using a polynomial function to the $\rhoc(T)$ data above 25~K and below 12~K, we can notice that $\rhoc(T)$ jumps by $\Delta\rhoc\approx 0.23~\Ohmcm$
, as indicated with the red arrow. This resistance drop reflects the decrease of the elastic scattering due to anion ordering.

Figures \ref{fig:raw}(b), (c) and (d) show the evolutions of the imaginary and real parts of the AC susceptibility ($\chiim(T)$ and $\chire(T)$ respectively) and of $\rhoc$ with the cooling rate.
All three quantities show that the sample cooled down at 0.02~K/min exhibits superconductivity below $\Tc(0.02~\text{K/min})\sim 1.3$~K. 
The superconducting features were progressively weakened upon increasing the cooling rate.
We comment that the small upturn in $\chiim(T)$ and $\chire(T)$ below 0.3~K is probably extrinsic, since such behavior was not observed in another sample (Sample~\#2; see Supplemental Material~\cite{NoteSupple}). Perhaps, the upturn could be attributed to small background signal originating from attached electrical leads and silver paint used for the resistivity measurements.

\begin{figure}[tbh]
\begin{center}
\includegraphics[width=8cm]{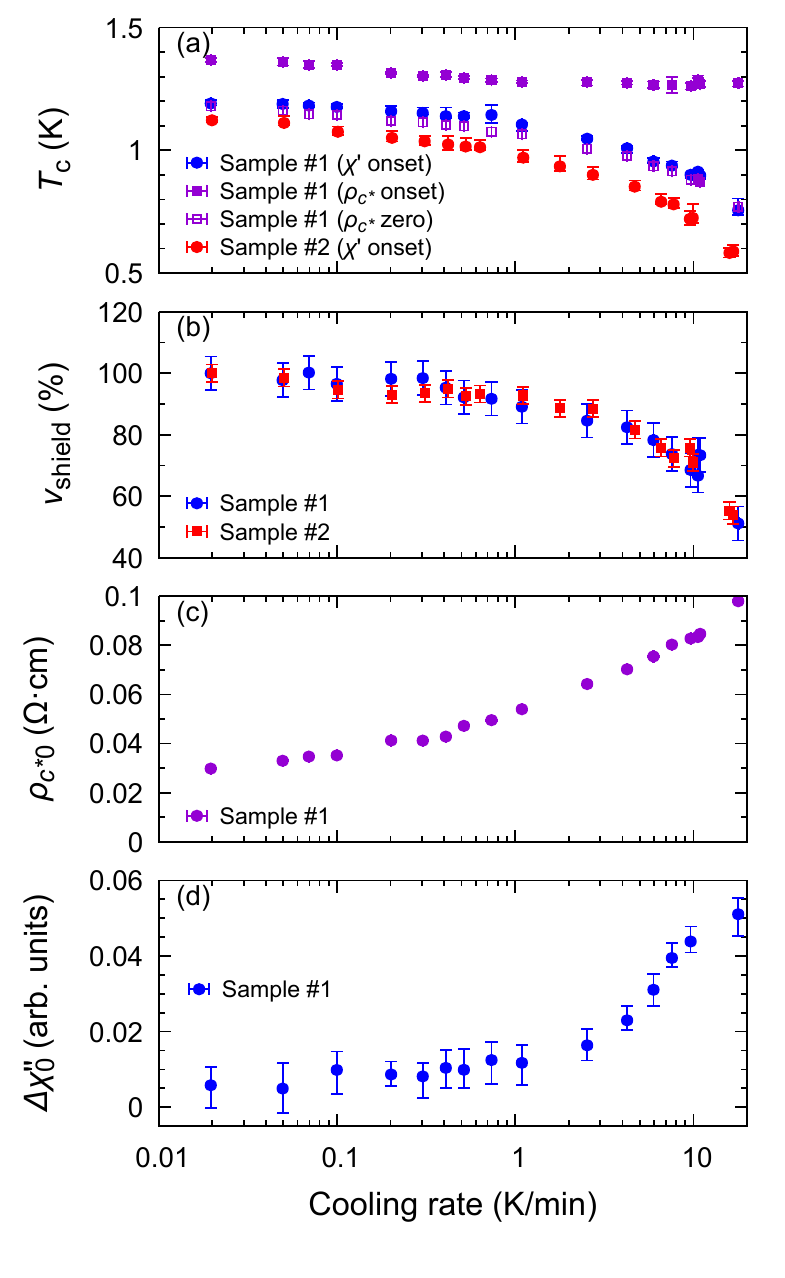}
\caption{(color online)
Cooling-rate dependence of superconducting properties for Sample \#1, and its comparison to that of Sample~\#2.
(a) Cooling-rate dependence of the various critical temperatures:
the blue circles  corresponds to $\Tcc$, the purple closed squares to $\Tcr$, and the purple open squares to $\Tcrz$.
The red circles indicate $\Tcc$ for Sample~\#2.
(b) Cooling-rate dependence of the relative shielded volume fraction $\VF$ for Sample~\#1 (blue) and \#2 (red).
(c) Cooling-rate dependence of the residual resistivity $\rho_{c^{\ast}0}$.
(d) Low temperature dissipation, as measured by $\chiim$ for Sample \#1. The $\chiim$ data have been averaged on the temperature range $0.1 \text{ K} < T < 0.3 \text{ K}$.
\label{fig:Tc-VF-rate}}
\end{center}
\end{figure}

Based on these raw data, the cooling-rate dependence of various superconducting properties can be examined:
\tcchi is defined  from the onset of $\chire(T)$
and \tcrho by the peak  in the second derivative of $\rhoc(T)$.
The zero-resistivity temperature $\Tcrz$ is the temperature where a linear extrapolation of the steep transition in $\rhoc(T)$ reaches zero.
We found that $\Tcrz$ is approximately equal to $\Tcc$.
The evolutions of the different $\Tc$'s with the cooling rate are given in Fig.~\ref{fig:Tc-VF-rate}(a). 
The small difference in $\Tcc$ for Samples \#1 and \#2 can be attributed to different ordering configurations which can exist even in samples of the same batch.
As  noticed in previous X-rays studies~\cite{Pouget12},  no perfect long range order of anions exists even in the slowest cooled samples. Let us moreover note that $\Tcc$ is not a thermodynamic quantity and, as we will later see, derives from a specific phase distribution within the sample.

Next, we focus on the change in the shielded volume fraction.
Since the samples cooled at rates smaller than 0.02~K/min are fully in the relaxed-state regime and are expected to display perfect diamagnetism at low temperatures, we will consider this cooling rate as a reference.
Figure~\ref{fig:Tc-VF-rate}(b) displays the cooling-rate dependence of the relative SC shielded volume fraction $\VF$, which is obtained from the zero temperature extrapolation of $\chire(T)$ divided by
its value for 0.02~K/min: $\VF \equiv |\chire(T \to 0)|/|\chire(T \to 0, \text{0.02~K/min})|$.

The zero temperature extrapolation of $\chiim$ averaged on the 0.1-0.3 K range is also given figure~\ref{fig:Tc-VF-rate}.d. We will discuss the meaning of this quantity in section \ref{granular behaviour}.

The normal-state residual resistivity $\rhocz$ presented in Fig.~\ref{fig:Tc-VF-rate}(c) is obtained by fitting the relation $\rhoc(T)=\rhocz + AT + BT^2$ to the data between 6 and 12~K (see the inset of Fig.~\ref{fig:raw}(d), as well as Fig.~\ref{fig:raw}(a)).
We chose this fitting range to avoid the influence of the downturn of $\rhoc(T)$, quite prominent below 5~K especially in fast cooled samples.
We note that the low-temperature downturn in $\rhoc(T)$ is ascribed to the sliding of  SDW fluctuations~\cite{Auban11} and will be examined in detail in a forthcoming publication.
Such a fitting procedure is justified by previous extensive investigations ascribing scattering against spin fluctuations as the source of the dominant linear temperature dependence of the resistivity~\cite{Doiron09,Doiron10,Sedeki12}.

Systematic decreases of all characteristic temperatures are observed with increasing cooling rates, as expected when disorder is increased.
However, although the decrease of \tcrho levels off at a value of $\approx 1.25-1.30$K above 1~K/min, \tcchi reveals a faster decrease above the same cooling rate.
The shielded volume fraction for $T\to 0$ steadily decreases with increasing cooling rate. Concomitantly, we notice an acceleration in the decrease of $\VF$ above 1~K/min.
No more than 50\% of the sample volume is shielded at 18~K/min.

\begin{figure}[tbh]
\begin{center}
\includegraphics[width=8cm]{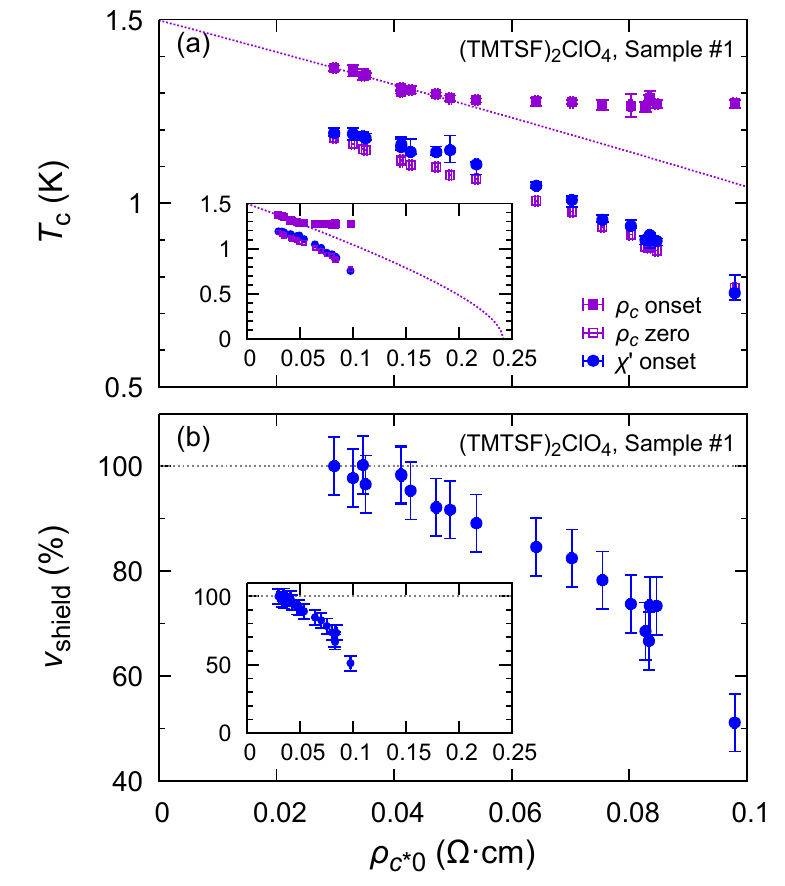}
\caption{(color online)
Dependence of $\Tc$'s and relative shielded volume fraction $\VF$ on $\rhocz$ for Sample~\#1.
(a) \tcrho as a function of $\rhocz$, compared with $\Tcrz$ and $\Tcc$.
The dotted curve presents results of fittings  of \tcrho with the AG formula (Eq.~\eqref{eq:digamma}).
The fitting has been performed in the range $0 < \rhocz < 0.05~\Ohmcm$.
The inset shows the same data in a wider range. 
(b) $\VF$ as a function of $\rhocz$.
The inset shows the same data in a wider range.
\label{fig:Tc-VF_vs_rho0}}
\end{center}
\end{figure}

Explaining the difference in behavior between \tcrho and \tcchi is the major issue of the present work and will underline the interpretation proposed in Sec.~\ref{granular behaviour}.

\section{Discussion}
\label{Discussion}

\subsection{Determination of the anion-ordered volume}
\label{sec:volume_determination}

So far, quantities such as $\Tc$, $\rhocz$ or $\VF$ were plotted against the cooling rate in Fig.~\ref{fig:Tc-VF-rate}.
Since the amount of impurities introduced by the cooling may depend on samples, as well as on the details of the cooling procedure, the physical picture is better conveyed when representing \tc and $\VF$ as functions of the residual resistivity, as shown in Figs.~\ref{fig:Tc-VF_vs_rho0}(a) and (b).
These figures reveal two regimes: below around $\rho_{c^{\star}0} \approx 0.05~\Ohmcm$ (corresponding to the cooling rate of $\approx 1$~K/min), $\Tc$, by all definitions, exhibits a steady decrease with increasing cooling rate whereas $\VF$ stays $\sim 100$\%. However, above this threshold, $\Tcr$ becomes nearly independent of the cooling rate, whereas $\VF$ clearly decreases.
This fact suggests a possible change in the sample  behavior.
Particularly, the substantial decrease in $\VF$ indicates that rapid cooling is no longer equivalent to creating local defects, and instead promotes the formation of clusters of disordered regions.

Expecting the picture of local defects to be no longer adequate at large cooling rates, it is important to derive the actual volume fraction in which bulk superconductivity develops since, in an inhomogeneous superconductor, it could be smaller than the shielded volume.

This volume fraction can be reached viewing the normal-state sample as a two-component composite conductor with   resistivities $\rhoMin$ and $\rhoMax$ for ordered and disordered regions respectively.
Such a problem of the mixture between two conductors of different conductivities is formally identical~\cite{Creyssels2017} to the effective dielectric permeability of a two-component dielectric medium treated by Landau and Lifshitz~\cite{Landau}.
The mixed conductor can thus be described at the lowest order by a three-dimensional (3D) effective medium theory with an effective conductivity $\sigma\subm{eff}=1/\rho\subm{eff}$,
which can be derived as
\begin{eqnarray}
{1/\rho_{\mathrm{eff}}=\left( p(1/\rhoMin)^{1/3}+(1-p)(1/\rhoMax)^{1/3}\right)^3}
\label{rhoeff}
\end{eqnarray}
where ${p}$ is the volume fraction of the anion-ordered domain.
The resistivity in the ordered region $\rhoMin$ is given by $\rhocz$ for 0.02~K/min, based on the assumption that the system at the lowest cooling rate provides $\rhoMin$.
The data in Fig.~\ref{fig:raw} lead to $\rhoMin= 0.030~\Ohmcm$.
For the disordered resistivity $\rhoMax$, the change in the elastic contribution $\Delta \rhoc$ across $\Ta$, coming from the frozen anion disorder, must be added to $\rhoMin$: i.e. $\rhoMax = \rhoMin + \Delta\rhoc = 0.26~\Ohmcm$~\cite{NoteModel}.

\begin{figure}[tbh]
\begin{center}
\includegraphics[width=8.5cm]{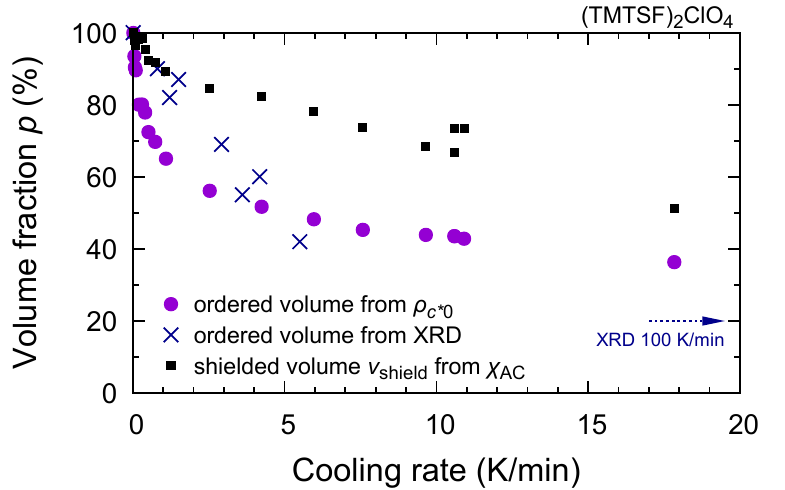}
\caption{(color online) Volume fraction of the anion-ordered domain derived from resistivity ($p$, obtained from the effective medium theory in this work; purple circles) and from high resolution X-rays ($\VX$, from Ref.~\cite{Pouget90}; blue crosses), compared with the shielded fraction evaluated from the AC susceptibility ($\VF$, black squares). 
The broken arrow shows the estimate for the ordered volume from X-ray measurements at 100~K/min~\cite{Kagoshima83}.
\label{Volumefractionvscooling}}
\end{center}
\end{figure}

\begin{figure}[tbh]
\begin{center}
\includegraphics[width=8.5cm]{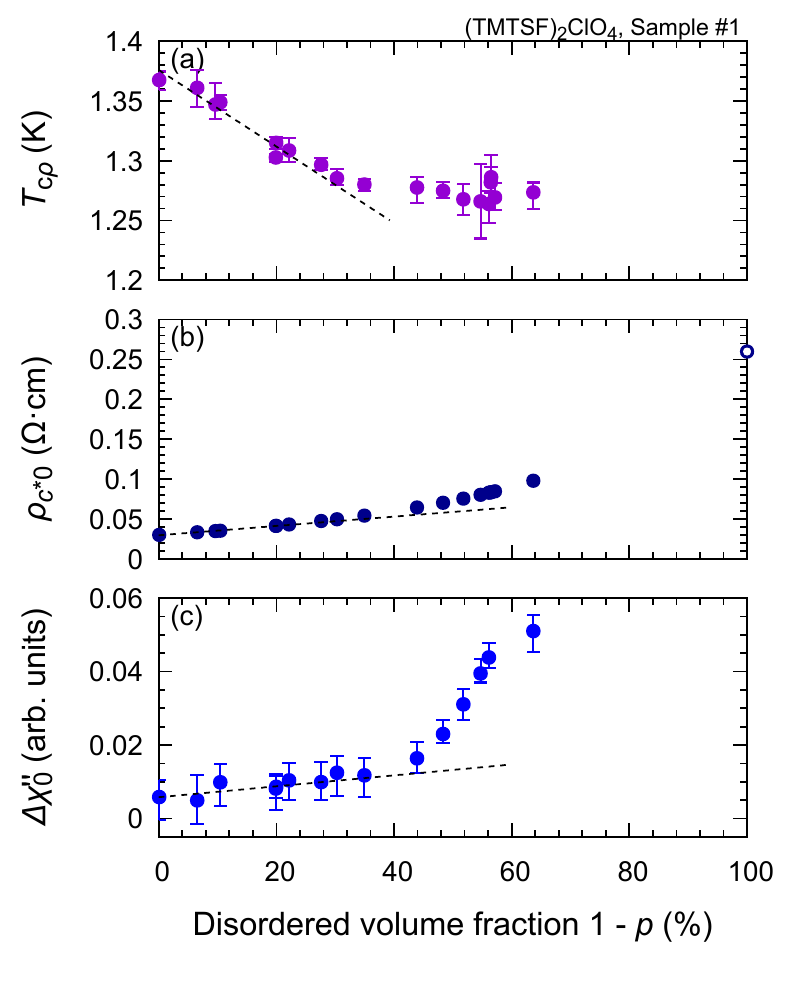}
\caption{(color online) 
(a) $\Tcr$, (b) $\rhocz$, and (c) low-temperature dissipation as measured by $\chiim$, plotted against the disordered volume fraction $1-p$ derived through Eq.~\eqref{rhoeff}. Deviation from the linear dependence (broken lines; fitted in the range $1-p < 20$\% in all panels) occurs when the first finite disordered cluster emerges. At complete disorder (not accessible before the onset of the SDW ground state above 18~K/min), the resistivity would amount to $\rhoMax = 0.26~\Ohmcm$, as indicated by the open circle. 
The $\chiim$ data in (c) have been averaged on the temperature range $0.1 \text{ K} < T < 0.3 \text{ K}$.
\label{rhoandTcversusordering}}
\end{center}
\end{figure}

Figure~\ref{Volumefractionvscooling} compares the ordered volume fraction $p$ obtained from Eq.~\ref{rhoeff} with the shielded volume $\VF$ and with the ordered volume fraction $\VX$ derived from X-ray measurements (Ref.~\citenum{Pouget90}).
Considering the uncertainties in the X-ray determination, there is a fair agreement between $\VX$ and $p$, particularly in the fast-cooling regime.
$p$ obtained at high cooling rates is also of the same order as $\VX$ at even higher disorder (broken arrow on Fig.~\ref{Volumefractionvscooling}).
Both $p$ and $\VX$ reflect the size of the ordered domains, which can reasonably be assumed to be SC.
In contrast, $\VF$ measures the shielded volume fraction.
Obviously, $p$ and $\VX$ are always smaller than $\VF$: in particular, at the highest cooling rate achieved in this study, $\VF$ is of about 50\%, whereas $p$ is reduced down to 36\%.
This fact can be seen as the signature of persistent shielding currents penetrating non-SC regions. The  derivation of the relation between the ordered volume fraction and the cooling rate (Fig.~\ref{Volumefractionvscooling}) enables us to show how $\Tcr$ and the residual resistivity evolve against the disordered volume fraction, see Fig.~\ref{rhoandTcversusordering}. This figure reveals the occurrence of a regime change for both $\Tcr$ and $\rhocz$ around 30\% disordered volume fraction. This feature will be studied more extensively in the following sections.

\subsection{Local defect regime}

Let us call the low cooling rate regime ($\leq 1$~K/min) ``Regime I'', where both \tcrho and \tcchi decrease linearly with $\rhocz$ (i.e. in the range $\rhocz \leq0.05~\Ohmcm$; see Fig.~\ref{fig:Tc-VF_vs_rho0}(a)).
In this regime, $\rhocz$ then is a good measure of disorder.
Hence the usual analysis for the dependence of \tc on non-magnetic disorder in $d$-wave superconductors~\cite{Sun95,Suzumura95} can be undertaken.

The Abrikosov-Gorkov theory~\cite{Abrikosov61} (AG) extended to non-magnetic impurities  in the case of non-$s$-wave superconductivity gives:
\begin{eqnarray}
\ln\left(\frac{\Tcz}{\Tc}\right)=\Psi\left(\frac{1}{2}+\frac{\alpha \Tcz}{2\pi \Tc}\right)-\Psi\left(\frac{1}{2}\right),
\label{eq:digamma}
\end{eqnarray}
where $\Psi(x)$ is the digamma function, $\alpha = \hbar /2 \tau k_{B}\Tcz$ the depairing parameter, $\tau$
the elastic scattering time, and $\Tcz$ the limit of $\Tc$ in the absence of any scattering (\textit{i.e.}, the limit of $\rhocz\rightarrow 0$).
The fit of $\Tc$ derived from Eq.~\eqref{eq:digamma} to the data is presented in Fig.~\ref{fig:Tc-VF_vs_rho0}(a).
Clearly, $\Tcr$ is well fitted in the slow-cooling (small $\rhocz$) regime.
This fact provides firm evidence that the superconductivity of \tmcns\ is of non-$s$-wave nature, and is consistent with previous works~\cite{Joo04,Joo05}.

Would it be driven by local defects only, the critical value of the residual resistivity for the suppression of superconductivity would read $\rhoczc=0.24~\Ohmcm$ from the fit to $\Tcr$ shown in the inset of Fig.~\ref{fig:Tc-VF_vs_rho0}(a).
This value is practically identical to the critical resistivity $\rhoczc=0.23~\Ohmcm$ obtained for the suppression of superconductivity on the  \cl-side of the \tmx solid solution due to the presence of non-magnetic \re \,anions~\cite{Jerome16}.
This gives additional confidence on the fact that, in Regime I, the destruction of superconductivity can be explained by the effect of local defects.

The nature of disorder in the present situation of an anion misorientation is quite different from the disorder in a solid solution~\cite{Joo05}.
The anion misorientation leads to disorder of the local electronic structures.
Hence, anions, to be active scattering centers, must form large enough clusters exhibiting a local electronic structure (the single sheet Fermi surface at $\pm k_F$) differing appreciably from the double-sheet Fermi-surface structure in the ordered regions~\cite{Alemany14}.
Such a condition requires the broadening of the Fermi surface due to a finite size of clusters to be kept at minimum, leading in turn to be at least 50 unit cells along the $a$ axis~\cite{Fermibroadening}.

The next task at hands is  the derivation of the elastic mean free path for the best ordered sample.
According to  magneto-transport data, we have $9\times 10^{-12}$~sec as the value for $\tau$ at the liquid helium temperature in the best-ordered situation~\cite{anisotropyoftau}, leading in turn to a corresponding mean free path $l_0$ along the $a$ axis~\cite{3Dregime} of 1620~nm, taking the $a$-axis Fermi velocity of $v_{\mathrm{F}a}= 1.8\times 10^{7}$~cm.s$^{-1}$.
Such an estimate for the mean free path is also in fair agreement with previous transport results, namely ($l_0=1400$~nm)~\cite{Tomic89}.
From the fitting of the AG theory to \tcrho in Regime I, the average mean free path at the border of Regime I is obtained as $l=972 $~nm at $1-p \sim 0.3$. Since $\Tcr$ is nearly constant beyond this regime, $l$ is reasonably assumed to be invariant, on average, for $1-p > 0.3$.

Furthermore, as already explained, the extrapolation of the AG formalism (inset of Fig \ref{fig:Tc-VF_vs_rho0}(a)) leads to the critical value $\rhoczc=0.24~\Ohmcm$, which is eight times larger than $\rhoMin$.
Thus, the critical mean free path $l\subm{c}$ corresponding to $\rhoczc$ is given by $l\subm{c} = l_0/8=202$~nm.
This is in good agreement with the critical mean free path $l\subm{c,d-wave}=\pi\xi_0\sim 220$~nm expected  for the suppression of $d$-wave superconductivity~\cite{Sun95}, given the measured SC coherence length along the $a$ axis, $\xi_{0a}=70$~nm~\cite{Murata87}.

We comment here on the width of the superconducting transition.
We claim that a broad distribution of the scattering length provides in turn a distribution of $\Tc$.
Indeed, it has been reported that, even in the ordered phase achieved by very slow cooling, in which \cl\ anions are ordered in one of their two possible ordering patterns, the domain size is largely distributed~\cite{Pouget12}. 
This is also supported by the fact that \tmcns\ ordinarily exhibits a broader superconducting transition than the pressure-induced superconductor \tmpns, which does not have any anion disorder because the PF\sub{6} anion is a centro-symmetric anion.
Moreover, as analyzed above using the AG model, scattering centers in Regime I are small-sized disordered clusters.
It is reasonable to assume that these clusters nucleate on the walls between ordered domains, because disorder ordinarily tends to develop in the vicinity of preexisting disorder, namely the domain walls in our case.
Then, the somewhat broad transition and the double-transition-like feature for our slowly cooled data (e.g. 0.02~K/min) can be attributed to such anion disorder remaining even under slow cooling. In addition, the sample can contain micro-cracks, which are often formed in organic crystals during cooling due to thermal-contraction strain, and such micro-cracks can also contribute to the broadness of the transition, by modulating $\Tc$ and/or by introducing additional weak-links within the sample. In Sample~\#1, we indeed observed a few small jumps in the temperature dependence of resistivity from room temperature to the lowest temperature. 
Such jumps indicate formation of a small number of micro-cracks.

\subsection{Evidences for a granular behaviour}
\label{granular behaviour}

Let us now call the higher cooling-rate regime ``Regime II'', for which $\rhocz>0.05~\Ohmcm$ and $1-p \geq 0.3$.
In this regime, $\Tcr$ is nearly constant ($T_{c\rho}=1.27\pm 0.03$~K), whereas the width of the resistive transition, measured by $\Tcr-\Tcrz$, increases as the cooling rate increases and the system becomes more and more disordered.
In this regime, the macroscopic SC coherence is weaker than what would be expected if superconductivity was controlled by local defects only.

In this section, we will demonstrate that this behavior can be interpreted as a consequence of granular superconductivity: SC puddles, consisting of an assembly of grains where ClO$_4$ anions are ordered along one of the ordering patterns (up-down-up-down \ldots\ or down-up-down-up \ldots; labeled as A or B, respectively)~\cite{Pouget12}, are distributed within a metallic background consisting of anion-disordered regions.

The progressive establishment of superconductivity in Regime II resembles in several respects the situations observed in the studies of superconductivity in 3D granular compounds~\cite{Oda79} and in 2D hybrid superconductor-normal metal-superconductor arrays~\cite{Deutscher80,Lobb83,Eley11}: superconductivity first appears in disconnected grains which couple via the proximity (or Josephson) effect at lower temperatures.

The system considered in this work differs from other granular materials~\cite{Oda79} insofar as, as we will see, the averaged puddle size is larger than the SC coherence length (the averaged coherence length within the $ab$ plane is $\xi_0=45$~nm; From the upper critical field measurements~\cite{Murata87}, $\xi$ were evaluated to be 70~nm, 30~nm, and 2~nm along the $a$, $b$, and $c$ axes, respectively; and the averaged $\xi_0$ mentioned here is evaluated using the geometrical mean $\sqrt{\xi_a\xi_b}$), but is smaller than the penetration depth.
Thus, bulk superconductivity is first established within each puddle while cooling.
Moreover, there is a strong coupling between neighboring puddles since the inter-puddles background (disordered regions) is metallic, albeit with a resistivity ($\rhoMax=0.26~\Ohmcm$) about 10 times larger than the puddle normal resistivity ($\rhoMin = 0.030~\Ohmcm$) (see Sec.~\ref{sec:volume_determination}). Let us emphasize that, in the present system, the grain size is tunable and solely governed by the cooling rate using a \emph{single} sample.

Thus, $\Tcr$ corresponds to the temperature at which the phenomenon of superconductivity begins to appear within each puddle.
It is governed by the puddle size $L$.
Although the puddles probably have a broad distribution in size, an estimate of the upper limit of $L$ can be derived from the AG approach.
Assuming that the electronic mean free path $l$ is limited by $L$, one then obtains $l\simeq L=970$~nm.
Such a large puddle size implies that they are composed of an equal weight of domains A and B, whose individual dimensions are of the order of 39~nm at 5~K/min, according to direct X-ray measurements~\cite{Pouget90}.
However, let us keep in mind that this estimated $L$ reflects the size of the largest puddles, which have the highest $\Tcr$.

A close look at the shape of the resistive transition shown in Fig.~\ref{fig:raw}(d) reveals that it takes place in two stages (particularly visible on the run at 18~K/min).
First, the resistance drops by a factor of about two before leveling off and giving rise to a second drop whose long tail ends at around $\Tcc$.
The temperature width of the first transition $\Delta T_1$ is analogous to the one measured in Regime~I ($\Delta T_1\simeq\left.\left(T_{c\rho}-T_{c\chi}\right)\right|_{\text{Regime I}}\simeq 0.2$~K).
It seems natural to ascribe the first drop to the onset of superconductivity in individual puddles, $\Delta T_1$ reflecting the distribution in the puddles size.
The second resistive drop can then be attributed to the progressive establishment of SC coherence over the whole sample: the SC puddles progressively couple through proximity effect to form SC clusters.
Let us also note that the value $1-p=0.3$, at which Regime II begins, is actually quite close to the percolation threshold for site percolation in a 3D cubic lattice~\cite{Kirkpatrick73,Stauffer94}.

Given the current structural knowledge on this system~\cite{Pouget90,Pouget12}, it is difficult to assess the shape and individual size of the SC puddles.
However, the nearly constant value of $\Delta T_1$ signifies that there is little change in the puddles size distribution in Regime II, at least before the inter-puddle coupling sets in.
This assumption is also supported by X-ray data~\cite{Pouget12}.
The decrease of the ordered volume fraction $p$ with the cooling rate therefore must be related to an increase of the average inter-puddle distance $d$.
The typical value of $d$ can be crudely approximated assuming spherical puddles forming a simple cubic lattice: $p = [(4\pi/3)(L/2)^3] / (L+d)^3$. For example, spheres of diameter $L=970$~nm lead to $d=130$~nm at $p=0.36$.

We here would like to comment on the distribution of the inter-puddle distance $d$. 
Although we assumed in the above analysis that $d$ is the same for all puddles, we do not think that the actual distribution of $d$ is so sharp. 
Indeed, when this is the case, the final resistance drop would be much sharper, as observed in e.g., Ref.~\cite{Eley2012.NaturePhys.8.59}. 
In the other limit, when there is a wide distribution in $d$, the final drop would be much broader and ill-defined~\cite{Allain2012.NatureMater.11.590}. 
Thus, the actual $d$ distribution in \tmcns\ is in a somewhat intermediate situation between the two limits.
Nevertheless, we continue to use the constant-$d$ assumption, which is the simplest model to capture the physics behind, in the model calculation and analysis below.

Once superconductivity of individual puddles is established, further cooling enables inter-puddles phase coherence via proximity effect through the normal metallic region in-between, in a somewhat similar manner to the behavior of Nb islands on Au~\cite{Eley11}. Macroscopic coherence is then reached when the first percolating path connects one end of the sample to the other.
For phase coherence to be established between neighboring puddles, the thermal energy needs to be smaller than the proximity effect energy~\cite{Likharev79,Lobb83,Eley11} for the local SNS junction:
$k\subm{B}T\lesssim E\subm{J}=\frac{\hbar I\subm{c}}{2e}$
where $E\subm{J}$ is the Josephson energy, and $I\subm{c}$ the Josephson current running between neighboring puddles.
Using the expression for $I\subm{c}$ in SNS junctions given by the Usadel equations in the diluted limit~\cite{Likharev79,Tinkham96}, one determines the temperature $\Tcc$ at which the macroscopic coherence is established:
\begin{equation}
    \Tcc = \frac{2\hbar}{\pi e^2k\subm{B}^2}\frac{\Delta_0^2( \Tcc )}{\Tcc}\frac{\mathcal{A}}{\rho\subm{N}}\frac{1}{\xi\subm{N}( \Tcc)}\mathrm{e}^{-[d/\xi_N( \Tcc)]},
\label{Tchi1}
\end{equation}
where $\rho\subm{N}$ is the resistivity of the metallic regions (N) and should be close to $\rhoMax$ in our case, $\mathcal{A}$ the averaged section of the junctions, $\Delta_0$ the SC gap in each puddle:
\begin{equation}
    \Delta_0^2(T)=\frac{8\pi^2}{7\zeta(3)}k\subm{B}^2 \Tcr^2 \left(1-\frac{T}{\Tcr}\right)
\label{eq:Delta}
\end{equation}
for $T$ close to $\Tcr$, and $\zeta(3)\simeq 1.202$.
In the Usadel equations, $\xi\subm{N}$, the normal-state coherence length in the N region, is determined by the dirty limit:
\begin{equation}
    \xi\subm{N}(T)=\sqrt{\frac{1}{3}\frac{\hbar v\subm{F}}{k\subm{B}T}l}.
\label{eq:xi_N}
\end{equation}
Notice that the clean-limit coherence length $\xi\subm{N}^0=\hbar v\subm{F}/kT$ is much larger than $\xi\subm{N}$, $d$ and $l \sim (\rhoMin/\rho\subm{N})l_0 \sim (\rhoMin/\rhoMax)l_0 \sim 190$~nm in our case (for example, $\xi\subm{N}^0=1400$~nm and $\xi\subm{N}=290$~nm at 1~K) and thus the dirty-limit treatment is justified.
Combining Eqs.~(\ref{Tchi1}) to (\ref{eq:xi_N}), the relation between $\Tcr$ and $\Tcc$ can be expressed as:
\begin{equation}
    \Tcc=\frac{\mathcal{C}}{\sqrt{\Tcc}}\left(1-\frac{\Tcc}{\Tcr}\right)\text{e}^{-[d(p)\sqrt{\Tcc}/\alpha]},
\label{Tchi}
\end{equation}
where $\mathcal{C}$ and $\alpha$ are constants. This equation can be numerically solved to analyze the data.

\begin{figure}[tb]
\begin{center}
\includegraphics[width=8.5cm]{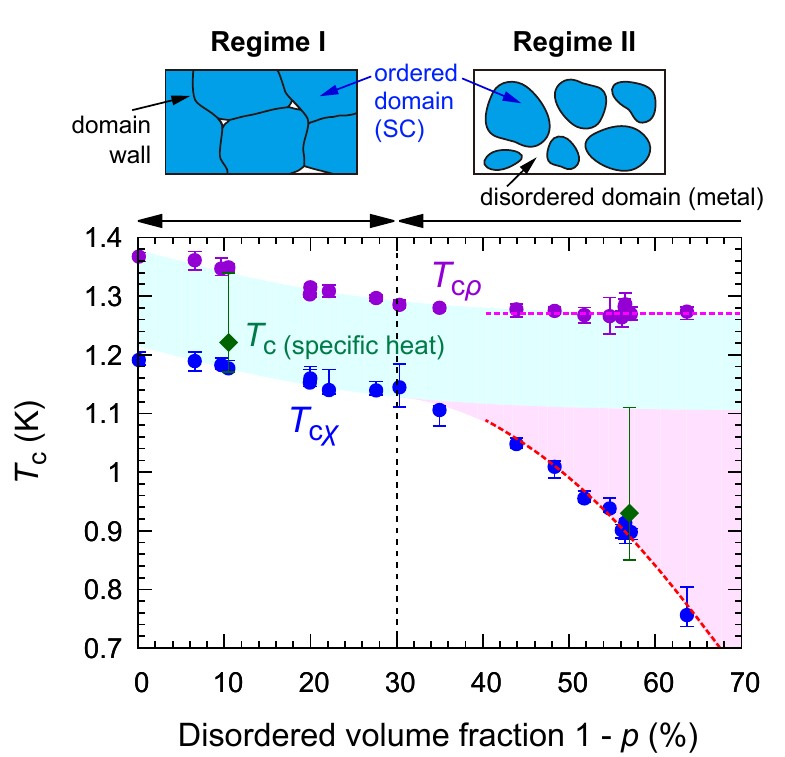}
\caption{(color online) Dependencies of $\Tcr$ (purple circles) and $\Tcc$ (blue circles) on the disordered volume fraction $1-p$ for Sample~\#1. Note that $\Tcc$ is practically equivalent to $\Tcrz$.
For comparison, we plot $\Tc$ from the specific heat study~\cite{Garoche82a}, with $1-p$ values estimated from the curves in Fig.~\ref{Volumefractionvscooling}.
For this data, the upper error bar corresponds to the onset of the transition, and the lower error bar to the peak in the electronic specific heat.
In the light blue region, superconductivity develops within individual puddles.
On the border between the light-blue and light-red regions, the proximity effect starts to induce SC coherence between puddles.
The large scale phase-coherent superconductivity establishes progressively in the light-red region as a result of the inter-puddle proximity effect coupling.
The broken curves indicate the region used for the analysis based on the proximity effect: $\Tcr$ is assumed to be constant (1.27~K) as shown with the magenta curve and $\Tcc$ is calculated by solving Eq.~\eqref{Tchi}.
The red curve is obtained by fitting the theory to the $\Tcc$ data in the range $1-p > 40$\%. The resulting fitting parameters are $\mathcal{C} = 5.2$~K\sps{3/2} and $\alpha = 100$~nm/K\sps{1/2}.
\label{diagramTcchi.pdf}}
\end{center}
\end{figure}

Let us now compare this expression to our experimental results in Fig.~\ref{diagramTcchi.pdf}.
Assuming $d(p)$ for the model of spherical puddles forming a simple cubic lattice, the observed evolution of $\Tcc$ with $p$ is compatible with that evaluated from Eq.~\eqref{Tchi} (the red curve in Fig.~\ref{diagramTcchi.pdf}).
The fitting parameters obtained are $\alpha=100$~nm/K$^{1/2}$ and $\mathcal{C} = 5.2$~K\sps{3/2}.
The value of $\alpha$ in turns gives $\xi\subm{N} = \alpha/\sqrt{T} =120$~nm at 0.75~K.
This value is in reasonable agreement what is expected in the disordered region~\cite{Xi_N}.
The fitting result $\mathcal{C} \simeq 5.2$ K$^{3/2}$ also compares favorably with its theoretical estimate:
assuming $\mathcal{A}\sim L^2\sim 9.4 \times 10^{-13}$~m\sps{2}, $v\subm{F} = 1.8\times 10^5$~m/s, $l=190$~nm, and $\Tcr=1.27$~K, we obtain $\mathcal{C}\spsm{theo} \simeq 49$~K$^{3/2}$.
Considering the randomness in the local distribution of puddles size, of their shape as well as of the inter-puddles distance that is not taken into consideration in our very crude model, $\mathcal{C}$ from experiment and theory agree reasonably with each other.

A second argument for this granular scenario comes from a previous specific-heat study
performed on a sample cooled at the rate of 10~K/min~\cite{Garoche82a} (corresponding to $1-p=57$\%;
see Fig.~\ref{Volumefractionvscooling}).
This study revealed a broad SC anomaly starting around 1.1~K and reaching a maximum around 0.85~K .
This  temperature for the maximum specific heat matches $\Tcc$, whereas the onset of the broad transition roughly corresponds to the border between the light-blue and light-red regions in Fig.~\ref{diagramTcchi.pdf}, namely the temperature domain where the pair condensation begins to be  significant thermodynamically speaking. Furthermore, specific heat data have revealed that the entropy involved in the SC pairing of this sample cooled at 10~K/min~\cite{Garoche82a} is only $50\pm10$\% of the  total electronic specific heat.  This feature  is in fair agreement with the $\approx 40$\% value for the bulk SC volume fraction derived  in  Fig.~\ref{Volumefractionvscooling}.
These agreements provide additional support for our scenario.
As demonstrated above, the proximity effect together with percolation represent a reasonable approach for understanding the development of the superconducting state in Regime~II.

Thirdly, another support for our scenario is provided by the AC susceptibility data. Indeed, it is striking that a magnetic signal is only perceptible below $\Tcc$, when the measured resistivity is only a small fraction of that of the normal state. This signifies that the part of the sample contributing to the initial resistivity drop is, in all likelihood, formed of small isolated SC puddles which size ($L\simeq 970$~nm) is smaller than or of the order of the SC penetration depth ($\lambda>$1000 nm~\cite{Le93,Greer03}), so that these are invisible to AC susceptibility measurements. The dissipation is then due to currents circulating in the normal (disordered) regions~\cite{comment_skin_depth}.

As the temperature is lowered, in-between $\Tcc$ and $\Tcr$, inter-grain coupling starts, so that SC clusters form, resulting in a first percolating path at $\Tcrz\simeq\Tcc$, but also in a sizable shielded volume.
Then $\chire$ begins to decrease, while $\chiim$ increases due to enhanced dissipation: as the SC volume expands, the current density in the remaining normal regions becomes also larger. More generally, as has been mentioned in the early studies of filamentary superconductors~\cite{Maxwell63,Strongin64,Oda79,Oda80}, the peak of $\chiim$ reflects a multi-phases compound. Here, it arises from a balance between the normal and SC regions: the peak occurs at a temperature $T_{\mathrm{peak}}$ when approximately half of the sample is shielded, as shown by the value of $\left.\chire\right|_{T\subm{peak}}\simeq-0.5$. This explains why, at higher cooling rates, $T\subm{peak}$ shifts to lower temperatures: at high cooling rates the initial density of SC puddles is lower, so that the coupling needs to be stronger, or the temperature lower, to achieve the same SC volume in the sample.

Finally, the $T\rightarrow0$ limit of $\chiim$ is a measure of the dissipation and hence an indirect probe of the disordered volume fraction. This is illustrated in Fig.~\ref{rhoandTcversusordering}(c). As can be seen, $\chiim(T\rightarrow0)$ is almost constant and close to 0 until $1-p\simeq 0.3$ (or equivalently, the cooling rate $\simeq$1 K/min; see Fig.~\ref{fig:Tc-VF-rate}(d)), before drastically increasing with the disordered volume fraction. The fact that $\chiim$ does not vanish even for $T\rightarrow0$ provides a yet another strong indication that normal regions persist in this limit.

Let us remark that this behavior is at variance with the behavior of cuprate high-$\Tc$ superconductors (HTSC). Indeed, in these compounds, the two peaks observed in the measurements of $\chiim$ have been attributed to the onset of intra- and inter-grain superconducting currents. Moreover, the initial drop in $\chire$ is steep in temperature, simultaneous with the resistive $\Tc$ and corresponds to the penetration of the field within each superconducting grain~\cite{Muller89,Nikolo95}. Finally, $\chire$ always goes to $-1$ in the low temperature limit. In our case, however, the shielding is not always total, the drop in $\chire$ is broad in temperature and only begins at $\Tcc$ where the resistivity measurements show that the percolation threshold is almost reached. In the present case, given the large size of the normal regions, the magnetic signal is more likely to be due to currents within the normal regions than to the penetration of currents in the SC regions, contrary to HTSC.

\section{Conclusion}

Our precise study  of the  normal and  superconducting electronic properties  of \tmc\ by simultaneous  transport and magnetic measurements, using a carefully controlled cooling procedure and covering cooling rates of four orders of magnitude, has revealed a crossover between a defect controlled $d$-wave and granular superconductivity in an inhomogeneous conductor with increasing cooling rate. 

At cooling rates up to about 1~K/min, non-magnetic disorder originates from small-volume clusters of disordered anions.
The clusters are randomly distributed and act as local scattering centers.
Superconductivity is suppressed by this disorder and the behavior of $\Tc$ versus the residual resistivity can be explained by the Abrikosov-Gorkov theory adapted to impure d-wave superconductors.
Above 1~K/min and up to 18~K/min, $\Tc$ for the resistivity onset remains independent of the cooling rate whereas $\Tc$ for the onset of the AC susceptibility signal keeps decreasing.
In this fast cooling-rate regime, the system behaves as a set of  randomly distributed SC (anion-ordered) puddles  embedded in a normal conducting background with disordered anions.
The  fraction of the sample volume comprising ordered anions has been derived using the theory for an effective medium composed of two regions with different residual resistivities.
Zero resistance and diamagnetic shielding  are achieved through  proximity effect links between the bulk SC puddles.
The volume fraction of bulk SC puddles is about 30\% smaller than the macroscopic screened sample volume measured by the real part of AC susceptibility.
This feature provides  an additional argument supporting the proximity effect picture in the fast-cooling regime.
We emphasize that the case of \tmc\ investigated in the present study is likely to be relevant  for other 1 or 2D molecular superconductors  when disorder arises either from molecular disorder or from the competition between other ground states~\cite{Senoussi06a,Muller09,Powell04}.
Similarly, the granular behavior at high disorder concentration can occur in other unconventional superconductors, particularly when the disorder is introduced chemically and form cluster-like structures.
Thus the present study provides a guideline and an insight toward future investigation of impurity effects in a wide variety of unconventional superconductors.

When considering the competition between SC and SDW in the two materials \tmpns\ and \tmcns, it becomes clear that the corresponding driving mechanisms are different.
For \tmpns\ under pressure, from the high-pressure SC side, SDW develops as a consequence of a decrease of the unnnesting term in the band structure (improving nesting in turn).
Only in a narrow pressure regime close to the critical pressure, can SC coexist with SDW \textit{albeit} in distinct domains in the sample.
In \tmcns\ on the other hand, the intermediate state consists of a mixture of anion-ordered puddles embedded in the anion-disordered background whose volume fraction depends on the cooling rate.

It would be particularly interesting to pursue this study to further understand how the electronic granularity arises from disorder. Comparing AC susceptibility measurements -- in zero and finite magnetic field -- in the different crystal orientations or STM measurements such as those performed on electronically inhomogeneous NbN thin films~\cite{Carbillet16} could for example provide information on the size of the SC puddles and the mechanism responsible for their formation.
Finally, let us note that the flexibility brought by the cooling procedure in tuning both the density and the dimensions of SC puddles and proximity effect bridges makes \tmc\ a promising candidate in which more advanced investigations of mesoscopic superconductivity could be undertaken.
One could, for instance, probe the quantum-critical metallic states that have been predicted to arise in such systems \cite{Spivak2008.PhysRevB.77.214523}, study how grain superconductivity and proximity effect each contribute to the specific heat as this issue has not yet been clearly settled \cite{Poran2017.NatureCommun.8.11464}, or study the localization-delocalization transition both in the three- and two-dimensional limits \cite{Gantmakher2008.PhysUsp.51.3}.

\begin{acknowledgements}

While this work was being carried out, the death of Professor Klaus Bechgaard had occurred and represents a great loss for the communities of chemistry and physics.
We have been  also very sad to hear about the death of Professor Lev Gorkov, a pioneer in the theory of superconductivity and a major contributor to the field of one dimensional conductors and organic superconductors.

S.~Y. acknowledges Y.~Maeno, Y.~Sugimoto, T.~Higuchi, K.~Ishida for experimental support and fruitful discussion.
D.~J. acknowledges fruitful exchanges with B.~Castaing, who made him aware of the Landau effective conductor model used in this work, and with J.~P.~Pouget for discussions on  X-rays data. 
We had several discussions with H.~Bouchiat, D.~Roditchev and T.~Cren on the topic of mesoscopic superconductivity.
This work has been supported by JSPS Grant-in-Aids KAKENHI JP26287078 and JP17H04848, as well as by JSPS Grant-in-Aids for Scientific Research on Innovative Areas ``Topological Materials Science''  (KAKENHI JP15H05852 and JP15K21717).

\end{acknowledgements}

\bibliography{ClO4_quenching_submit.bib}

\begin{thebibliography}{106}
\bibitem [{Not(){\natexlab{a}}}]{NoteRhoC}%
  \BibitemOpen
  \bibinfo {note} {The reason why we chose to measure $\rhoc$ is that the
  reproducibility of the data among different runs is especially important for
  the present investigation: Measurements of transport along the $a$ or $b$
  axes are often perturbed by sample cracks occuring during cooling. In
  contrast, $\rhoc$ is known to be less influenced by such cracks even after
  multiple cooling processes~\cite{Cooper86}. In addition, magnetoresistance
  measurements~\cite{Cooper86} have shown that the band theory should apply
  below 10~K as also supported by the existence of a Drude edge along the
  $\cstar$ axis~\cite{Henderson99}.}\BibitemShut {Stop}%
\bibitem [{Not(){\natexlab{c}}}]{NoteModel}%
  \BibitemOpen
  \bibinfo {note} {The model which is used here intends to establish a
  relationship between the cooling rate and the volume fraction in order to
  link experimental results to the theory of percolation to be mentioned later.
  It may be oversimplified as it neglects the weak volume dependence of the
  residual resistivity of ordered regions in Regime I discussed in Sec
  IVB.}\BibitemShut {Stop}%
\bibitem [{ani()}]{anisotropyoftau}%
  \BibitemOpen
  \bibinfo {note} {According to the transport data displaying $\Tc$ vs
  $\rho_{c^{\ast}0}$, the critical scattering rate,
  $\tau\subm{c}^{-1}=0.88k\subm{B}T\subm{c0}$ and $T\subm{c0}=1.5$~K, provides
  $\tau\subm{c} = 5.8\times10^{-12}$~sec leading in turn to $\tau_0 =
  4.6\times10^{-11}$~sec for the lowest cooling rate. This latter elastic
  lifetime is admittedly longer than values derived from the analysis of
  angular magnetoresistance (AMRO) experiments in \tmc~\cite{Sugawara06}. Such
  a situation has already been encountered and commented in a previous
  study~\cite{Yonezawa08a}. It has been attributed to the anisotropy in
  impurity scattering which may weaken the suppression of $\Tc$ while a
  $k$-averaged shorter $\tau$ should appear fully in the theory of angular
  magnetoresistance}\BibitemShut {NoStop}%
\bibitem [{3Dr()}]{3Dregime}%
  \BibitemOpen
  \bibinfo {note} {In this rest analysis, we use the longitudinal scattering
  rate $1/\tau_0$ as a measure for the elastic disorder governing $\Tc$
  although the corresponding mean free path is known to be highly
  anisotropic~\cite{Cooper86}. Previous comparative studies of resistivities
  along the $a$ and $c^{\ast}$ directions have shown~\cite{Auban11a,Cooper86}
  that both resistivities exhibit similar temperature dependences up to 30~K in
  \tmcns. This behaviour provides an additional confirmation for the existence
  of a 3D coherent regime at low temperature (albeit the non-Fermi-liquid-like
  temperature dependence of the resistivity).}\BibitemShut {Stop}%
\bibitem [{com()}]{comment_skin_depth}%
  \BibitemOpen
  \bibinfo {note} {The sample dimensions are smaller than the skin depth
  $\delta$ at the considered frequencies: $\delta\simeq2$~cm.}\BibitemShut
  {Stop}%
\end{thebibliography}


\onecolumngrid

\clearpage

\renewcommand{\thesection}{S\Roman{section}}
\setcounter{section}{0}
\renewcommand{\theequation}{S\arabic{equation}}
\setcounter{equation}{0}
\renewcommand{\thefigure}{S\arabic{figure}}
\setcounter{figure}{0}
\renewcommand{\thetable}{S\arabic{table}}
\setcounter{table}{0}
\renewcommand{\thepage}{S\arabic{page}}
\setcounter{page}{1}

\ 

\vspace{0.3cm}

\begin{center}
{\large
\textmd{Supplemental Material for}\\[0.2cm]
\textbf{Crossover from impurity-controlled to granular superconductivity in (TMTSF)$_{\bm{2}}$ClO$_{\bm{4}}$}
}

\vspace{0.3cm}

Shingo~Yonezawa$^{1\ast}$, Claire A Marrache-Kikuchi$^2$, Klaus~Bechgaard$^3$, Denis~J\'erome$^4$

\vspace{0.15cm}

\noindent
${}^1$ \textit{\small Department of Physics, Graduate School of Science,  Kyoto University, Kyoto 606-8502, Japan}

\noindent
${}^2$ \textit{\small  CSNSM, Univ. Paris-Sud, CNRS/IN2P3, 91405 Orsay, France}

\noindent
${}^3$ \textit{\small  Department~of~Chemistry, Oersted~Institute, Universitetsparken 5, 2100 Copenhagen, Denmark}

\noindent
${}^4$ \textit{\small  Laboratoire de Physique des Solides (UMR 8502), Univ. Paris-Sud, 91405 Orsay, France}

\noindent
${}^{\ast}$ yonezawa@scphys.kyoto-u.ac.jp

\end{center}

\vspace{1.cm}

\twocolumngrid

\section{Details of the experimental configuration}

In this study, we performed two sets of experiments with different experimental setups. 

\begin{figure}[tb]
\begin{center}
\includegraphics[width=8cm]{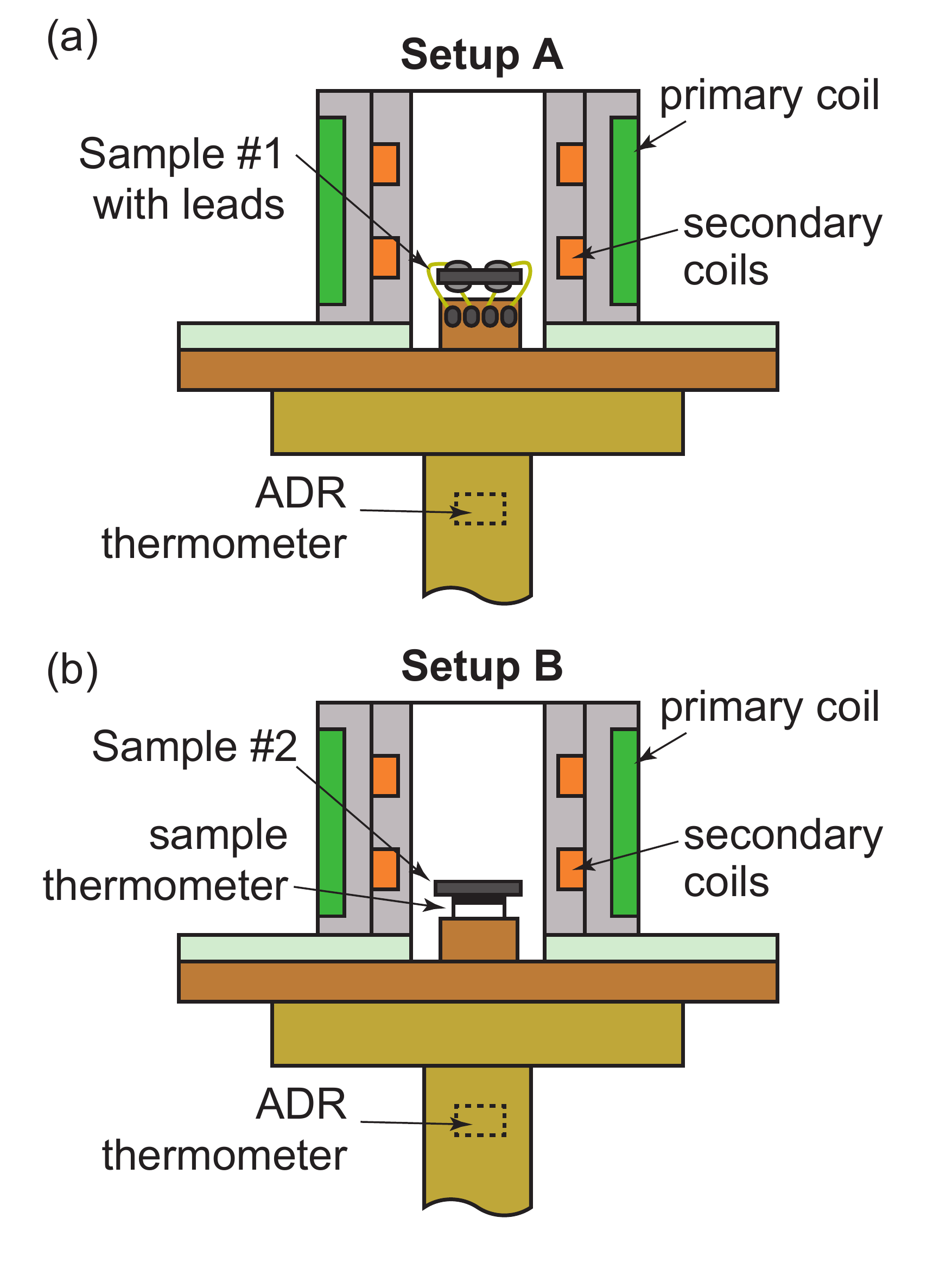}
\caption{(color online)
Schematic description of our experimental setups.
\label{fig:setups}}
\end{center}
\end{figure}

For Setup A, Sample~\#1, whose results are mainly discussed in the main text, was fixed onto a sample stage inside the susceptometer, as shown in Fig.~\ref{fig:setups}(a).
This sample accompanied four electrodes for $\cstar$-axis resistivity measurements attached onto the $ab$ faces of the crystal.
This setup enabled simultaneous measurements of the resistivity and AC susceptibility of the sample, whereas we had to rely on the built-in thermometer of the ADR option because of the limitation in the available electrical leads.
Nevertheless, during the experiment using Setup B, we have checked that the difference between the sample temperature and the temperature of the ADR option 
was less than 1~K even during the cooling procedure for the fastest cooling rate (16.7~K/min) achieved in this experiment.
The temperature difference during measurement, which was performed during warming after reaching 0.1~K with the adiabatic demagnetization cooling method (see the next section for the details of the experimental sequence), was up to 0.02~K at the lowest temperature (below 0.2~K) but was less than 0.01~K in the other measurement temperature ranges.

For the Setup B, Sample~\#2 was fixed directly onto a small calibrated thermometer (ruthenium-oxide thick-film resistor) placed inside the susceptometer, as schematically shown in Fig.~\ref{fig:setups}(b).
With this setup, we can measure the sample temperature and the actual cooling rate more accurately.

For both setups, we avoided using magnetic and superconducting materials (such as gold-plated parts with nickel buffer and solder) inside and around the susceptometer coil, to minimize artificial background signals.
For example, we used silver paste or silver epoxy to connect electrical leads.

\section{Typical experimental process}

\begin{figure}[tb]
\begin{center}
\includegraphics[width=8cm]{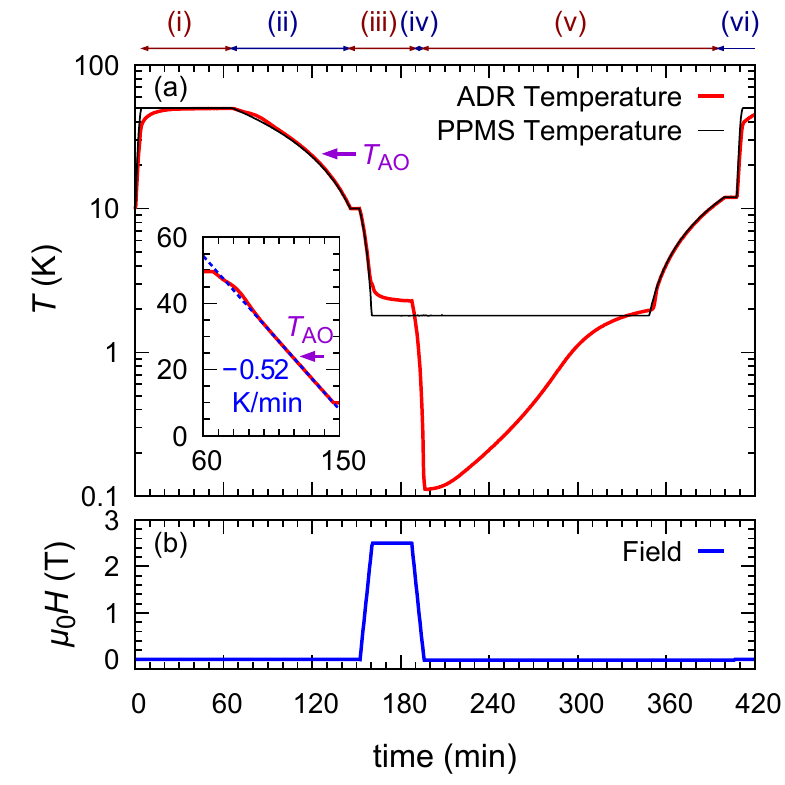}
\caption{
(color online)
Typical cooling and measurement process. 
(a) Time dependence of the system temperature (PPMS temperature; black curve) and the temperature of the adiabatic demagnetization refrigeration (ADR) option (ADR temperature; red curve) during the measurement of the $-0.52$~K/min data of Sample~\#1.  
The inset shows the ADR temperature in the step (ii) with the linear vertical scale, together with the result of a linear fitting (blue broken line) performed in the range $20 < T < 30$~K.
(b) Time dependence of the applied magnetic field. 
See the text for the explanation of the steps (i)-(vi).
\label{fig:sequence}
}
\end{center}
\end{figure}

In Fig.~\ref{fig:sequence}, we present the time evolutions of the temperatures and the magnetic field for one set of cooling and measurement for the setup A. 

The experiment was performed along the following steps: 
(i) The Physical Properties Measurement System (PPMS) system was heated to and kept at 50~K for 1 hour in order to fully randomize the anions' direction.
(ii) The system was cooled down to 10~K with the targeted cooling rate ($-0.5$~K/min for the case of Fig.~\ref{fig:sequence}). 
It is this cooling rate that is discussed in the main text.
(iii) The system was further cooled to 1.8~K and a magnetic field of 2.5~T was applied. After reaching 1.8~K, the sample space was evacuated if necessary to achieve an adiabatic condition. Note that, below around 10~K, we can use any cooling rate because changes of the cooling rate do not affect the degree of anion disorder, which is completely frozen below this temperature range.
(iv) After the ADR temperature was settled, the magnetic field was turned off to cool the sample to 0.1~K via the adiabatic demagnetization refrigeration. 
Here, we actually set the field to a slightly negative value (typically $-150$~Oe) to cancel out the remnant field.
(v) AC susceptibility and resistivity were measured during warming.
(vi) The system was again heated to 50~K for the next set of cooling and measurement.

\section{AC susceptibility results for Sample~\#2}

In this section, we present AC susceptibility results for another sample, Sample~\#2.
For this sample, only the susceptibility, not resistivity, was measured.

\begin{figure}[tb]
\begin{center}
\includegraphics[width=8cm]{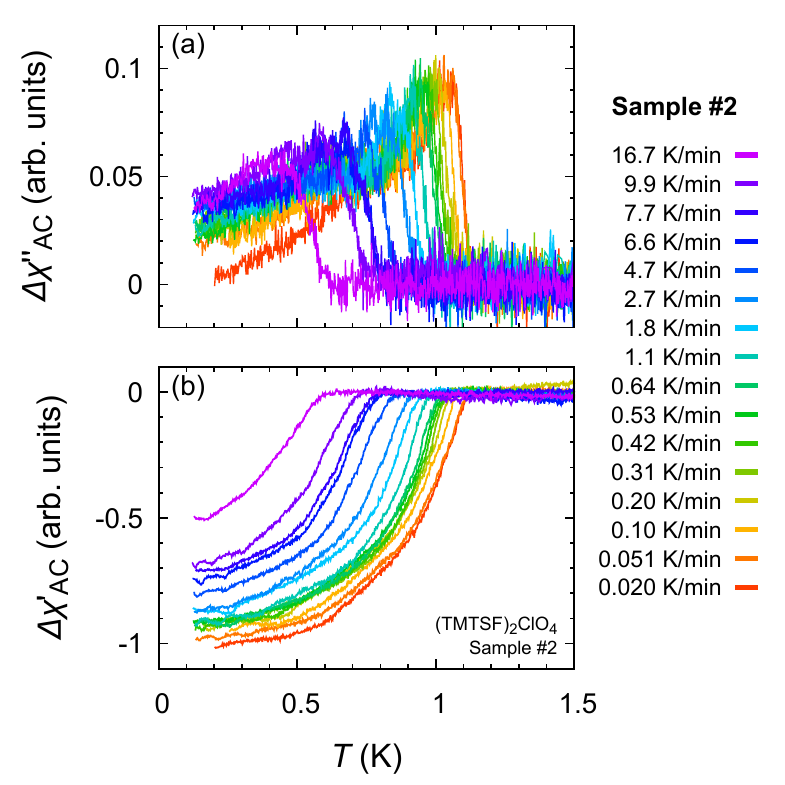}
\caption{(color online)
Temperature dependence of the AC susceptibility of Sample~\#2 for various cooling rates, measured with 887~Hz and $\sim 0.17$~Oe-rms AC magnetic field.
\label{fig:chi-AC-sample2}}
\end{center}
\end{figure}

Figure~\ref{fig:chi-AC-sample2} presents the temperature dependence of the imaginary and real parts of the AC susceptibility ($\chiim$ and $\chire$, respectively) of Sample~\#2 measured after cooling across the anion ordering temperature $\Ta = 24$~K.
For the real part, the overall evolution of the $\chire(T)$ curve by increase of the cooling rate is quite consistent with that observed for Sample~\#1 (see Fig.~1(c) the main text):
the onset temperature $\Tcc$ as well as the shielded volume fraction $\VF$ is suppressed by increase of the disorder.
Indeed, the cooling-rate dependences of $\Tcc$ and $\VF$ both behave similarly to those of Sample~\#1 (see Figs.~2(a) and (b)).
A small downshift of $\Tcc$ of Sample~\#2 probably due to chemical impurities, different anion-order domain configuration, and/or small micro-cracks in the sample occurred during cooling from room temperature.

\begin{figure}[tb]
\begin{center}
\includegraphics[width=8cm]{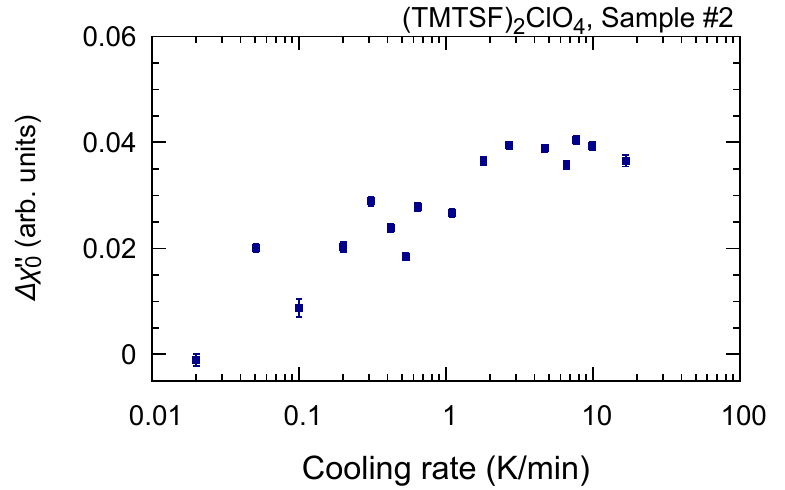}
\caption{(color online)
Cooling-rate dependence of the extrapolation of $\chiim$ to $T\to 0$ of Sample~\#2.
\label{fig:chi-im-vs-rate}}
\end{center}
\end{figure}

The imaginary part $\chiim$ of Sample~\#2 behaves slightly differently to that of Sample~\#1.
Firstly, for the slow cooling, the peak in $\chiim$ at $\Tcc$ of Sample~\#2 is sharper than that of Sample~\#1.
This fact suggests that the $\Tc$ distribution that exists even for an ideal case where anion disorder is completely absent is narrower in this sample.
Secondly, in Sample~\#2, $\chiim$ remains finite even around 0.3 K and for slow cooling, while $\chiim$ is nearly zero below around half of $\Tcc$ for Sample~\#1.
This difference might be attributable to weak links due to micro cracks formed during cooling. 
Such weak links would result in a wider distribution of the Josephson coupling between adjacent superconducting islands in Sample~\#2 than in Sample~\#1.
Because of this difference, we used a linear extrapolation of $\chiim$ to $T\to 0$ as the representative low-temperature value of the energy dissipation. 
This extrapolated $\chiim$ is plotted in Fig.~\ref{fig:chi-im-vs-rate} as a function of the cooling rate.
In contrast to the Sample-\#1 data shown in Fig.~2(d), only a somewhat smeared upturn at around 1~K/min is observed for Sample~\#2.
This is consistent with a larger distribution of the coupling constants between the superconducting islands.

%

\end{document}